\documentclass[apj,twocolappendix,numberedappendix]{emulateapj}

\usepackage{times}
\usepackage{graphicx}
\usepackage{amssymb}
\usepackage{amsmath}
\usepackage{pifont}
\usepackage{graphics}
\usepackage{xcolor}
\usepackage{epsfig}
\usepackage[hidelinks]{hyperref}
\usepackage{amssymb}
\usepackage{courier}

\newcommand{\degrees}{\ensuremath{^{\circ}}}
\newcommand{\chisq}{\ensuremath{\chi^{\,2}}}
\newcommand{\bjdtdb}{\ensuremath{\rm {BJD_{TDB}}}}

\usepackage[normalem]{ulem}
\usepackage{color}

\begin{document}

\title{AstroImageJ: Image Processing and Photometric Extraction for Ultra-Precise Astronomical Light Curves}

\author{
Karen A.\ Collins\altaffilmark{1,2,3}, 
John F.\ Kielkopf\altaffilmark{\,3},
Keivan G.\ Stassun\altaffilmark{1,2}, and
Frederic V.\ Hessman\altaffilmark{\,4}
}
\altaffiltext{1}{Department of Physics \& Astronomy, Vanderbilt University, Nashville, TN 37235, USA, karenacollins@outlook.com}
\altaffiltext{2}{Department of Physics, Fisk University, Nashville, TN 37208, USA}
\altaffiltext{3}{Department of Physics \& Astronomy, University of Louisville, Louisville, KY 40292, USA}
\altaffiltext{4}{Inst. f. Astrophysik, Georg-August-Universit\"{a}t G\"{o}ttingen}

\begin{abstract}
ImageJ is a graphical user interface (GUI) driven, public domain, Java-based, software package for general image processing traditionally used mainly in life sciences fields. The image processing capabilities of ImageJ are useful and extendable to other scientific fields. Here we present AstroImageJ (AIJ), which provides an astronomy specific image display environment and tools for astronomy specific image calibration and data reduction. Although AIJ maintains the general purpose image processing capabilities of ImageJ, AIJ is streamlined for time-series differential photometry, light curve detrending and fitting, and light curve plotting, especially for applications requiring ultra-precise light curves (e.g., exoplanet transits). AIJ reads and writes standard FITS files, as well as other common image formats, provides FITS header viewing and editing, and is World Coordinate System (WCS) aware, including an automated interface to the astrometry.net web portal for plate solving images. AIJ provides research grade image calibration and analysis tools with a GUI driven approach, and easily installed cross-platform compatibility.  It enables new users, even at the level of undergraduate student, high school student, or amateur astronomer, to quickly start processing, modeling, and plotting astronomical image data with one tightly integrated software package. 
\end{abstract}


\section{Introduction}
\label{sec:intro}

In many areas of astronomy, there is a need for image processing and analysis capabilities and light curve extraction. One such general purpose environment is IRAF \citep{Tody:1986,Tody:1993}\footnote{IRAF is distributed by the National Optical Astronomy Observatories, which are operated by the Association of Universities for Research in Astronomy, Inc., under cooperative agreement with the National Science Foundation.}. However, especially for ultra-precise photometry in fast-paced areas of research such as exoplanet transits and microlensing, there remains a need for a general, off-the-shelf integrated analysis environment that is at once sophisticated yet easy to use. Indeed, citizen science and professional-amateur collaborations increasingly require robust tools that can deliver research-grade results while enabling broad usability.

Here we present AstroImageJ (AIJ), an astronomical image analysis software package that is based on ImageJ (IJ; \citealt{Rasband:1997}), but includes customizations to the base IJ code and a packaged set of software plugins that provide an astronomy specific image display environment and tools for astronomy specific data reduction, analysis, modeling, and plotting. AIJ and IJ are public domain, open source, Java programs inspired by NIH Image for the Macintosh computer. Some AIJ capabilities were derived from the \textit{Astronomy} plugins package\footnote{\url{http://www.astro.physik.uni-goettingen.de/~hessman/ImageJ/Astronomy/}}. Some astronomical algorithms are based on code from \textit{JSkyCalc} written by John Thorstensen of Dartmouth College. Because AIJ is Java code, the package is compatible with computers running Apple OS X, Microsoft Windows, and the Linux operating system (OS).

AIJ is a general purpose astronomical image processing tool, plus it provides interfaces to streamline the interactive processing of image sequences. The current release (version 3.1.0) includes the following features and capabilities, where (I) indicates a feature provided by the underlying ImageJ platform, (I+) indicates an ImageJ feature that has been improved, (A+) indicates a feature based on the \textit{Astronomy} plugins package, but with significant new capabilities, and (N) indicates a new feature that is available in AIJ, but not available in ImageJ or the \textit{Astronomy} plugins package:

\begin{itemize}
  \item (N) Interactive astronomical image display supporting multiple image stacks with fast image zooming and panning, high-precision contrast adjustment, and pixel data display similar to \textit{SAOImage DS9} \citep{DS9:2000}\footnote{SAOImage DS9 has been made possible by funding from the Chandra X-ray Science Center (CXC) and the High Energy Astrophysics Science Archive Center (HEASARC).}
  \item (N) Live mouse pointer photometer
  \item (N) Sky orientation of image and pixel scale display in a non-destructive image overlay
  \item (A+) Reads and writes FITS images with standard headers, as well as most other common image formats (e.g. tiff, jpeg, png, etc.)
  \item (N) Data Processor facility for image calibration including bias, dark, flat, and non-linearity correction with an option to run in real-time during observations
  \item (A+) Interactive time-series multi-aperture differential photometry with detrend parameter extraction, and an option to run in real-time during observations
  \item (N) Photometric uncertainty calculations including source and sky Poisson noise, dark current, detector readout noise, and quantization noise, with automatic propagation of single-aperture uncertainties through differential photometry, normalization, and magnitude calculations 
  \item (N) Comparison star ensemble changes without re-running differential photometry 
  \item (N) Interactive multi-curve plotting streamlined for display of light curves 
  \item (N) Interactive light curve fitting with simultaneous detrending 
  \item (N) Plate solving and addition of WCS headers to images seamlessly using the Astrometry.net web interface
  \item (N) Time and coordinate conversion with capability to update/enhance FITS header content (airmass, \bjdtdb, etc.)
  \item (A+) Point and click radial profile (i.e.\ seeing profile) plots
  \item (N) FITS header viewing and editing
  \item (N) Astronomical coordinate display for images with WCS 
  \item (N) Object identification via an embedded SIMBAD interface 
  \item (A+) Image alignment using WCS headers or apertures to correlate stars
  \item (N) Non-destructive object annotations/labels using FITS header keywords
  \item (I) Mathematical operations of one image on another or an image stack, and mathematical and logical operations on single images or image stacks
  \item (I+) Color image creation
  \item (N) Optionally enter reference star apparent magnitudes to calculate target star magnitudes automatically
  \item (N) Optionally create Minor Planet Center (MPC) format for direct submission of data to the MPC 
\end{itemize}

AIJ is currently used by most of the $\sim 30$ member Kilo-degree Extremely Little Telescope (KELT; \citealt{Pepper:2003,Pepper:2007}) transit survey photometric follow-up team, 
so far resulting in 10 planets published
\citep{Siverd:2012,Beatty:2012,Pepper:2013,Collins:2014,Bieryla:2015,Fulton:2015,Eastman:2016,Rodriguez:2016,Kuhn:2016}, 
and at least 8 more in press or in preparation as of this writing. 
AIJ users on the team include amateur astronomers, undergraduate and graduate students, and professional astronomers. AIJ is also used by the KELT science team to optimize the precision of, and determine the best detrending parameters for, all follow-up light curves that are included in the analysis of new planet discoveries. AIJ is deployed in multiple undergraduate university teaching labs and is also used to teach exoplanet transit analysis to high school students. We and the KELT follow-up team have verified the accuracy of AIJ against a number of traditional scientific and commercial photometric extraction packages, including IRAF, IDL\footnote{http://www.harrisgeospatial.com; IDL is a product of Exelis Visual Information Solutions, Inc., a subsidiary of Harris Corporation.}\textsuperscript{,}\footnote{http://idlastro.gsfc.nasa.gov}, and MaxIm DL\footnote{http://www.cyanogen.com}. The IRAF and IDL photometric capabilities were adapted from DAOPHOT \citep{Stetson:1987}. We do not track the number of AIJ downloads, but we estimate that there are several hundred active AIJ users based on AIJ user forum\footnote{\label{userforum}\url{http://astroimagej.1065399.n5.nabble.com/}} statistics.

AIJ's ultra-precise photometric capabilities are demonstrated by \citet{Collins:2017a}, where they achieved a RMS of 183 and 255 parts per million for the transit model residuals of the combined and five minute binned ground-based light curves of WASP-12b and Qatar-1b, respectively, and transit timing residuals from a linear ephemeris of less than $\sim30$~s. These results are enabled by a multi-star photometer that allows fixed or variable radius apertures, a variety of options to calculate sky-background, including sky-background star rejection, and high precision centroid capabilities, including the ability to properly centroid on defocused stars. In addition, AIJ's interactive GUIs and tightly coupled extraction of differential photometry and detrend parameters, light curve plotting, comparison star ensemble manipulation, and simultaneous fitting of the data to a transit model and detrending parameters, enable the user to quickly optimize detrended and fitted light curve precision. 

For example, stars can be added to or removed from the comparison ensemble (without re-running photometry) and detrending parameters can be changed instantly by clicking to enable or disable each one. When a change is made, the light curve and fitted model plots are automatically updated and statistical values indicating the goodness of the model fit, such as RMS and the Bayesian Information Criterion, are instantly updated. These interactive features enable a user to quickly determine the best aperture settings, comparison ensemble, and detrend parameter set. 

Finally, if AIJ is operated in ``real-time'' mode during time-series observations, the images are calibrated, photometry is extracted, and data are plotted, detrended, and model fitted automatically as images are written to the local system's hard disk by any camera control software package. This capability works independent of (and does not interfere with) an observatory's telescope and camera control software and allows the user to explore exposure time, defocus, aperture settings, and comparison star ensemble to ensure high-precision photometric results in the final post-processed data. 

The following sections provide an introduction to the astronomy specific capabilities of AIJ. An enhanced version of this work \citep{Collins:2017c}, the AIJ User Guide, installation packages, and installation instructions are available for download at the AIJ website\footnote{\url{http://www.astro.louisville.edu/software/astroimagej}}. Most of the AIJ user interface panels include ``tool-tip'' help messages that optionally pop up when the mouse pointer is positioned over an item in the display for more than a second. An AIJ user forum\textsuperscript{\ref{userforum}} is available to facilitate shared support for the software. AIJ inherits all of the basic image manipulation and analysis functionality from IJ. The IJ website\footnote{\url{http://imagej.nih.gov/ij/}} provides detailed user guides and descriptions of its functionality.

\section{AIJ Overview and Basic Capabilities}

\subsection{Toolbar}

\begin{figure*}
\begin{center}
\resizebox{\textwidth}{!}{
\includegraphics*{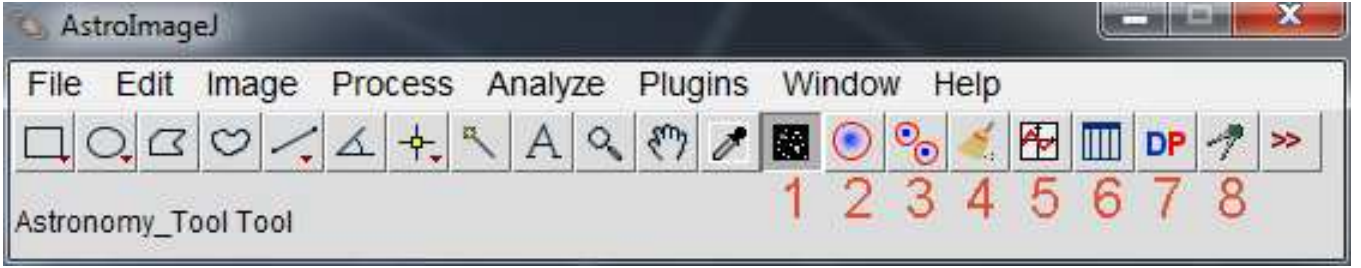} }
\caption[AIJ Toolbar]{The AIJ Toolbar. The icon shown depressed and labeled as 1 indicates that AIJ is in astronomy mode. In this mode, all images open into the Astronomical Image Display mode discussed in \S \ref{sec:aijdisplay}. Icon 2 starts the single aperture photometry mode discussed in \S \ref{sec:singleaperture}. A double-click on Icon 2 opens the \textit{Aperture Photometry Settings} panel discussed in Appendix \ref{sec:photset}. Icon 3 starts the Multi-Aperture photometer module discussed in \S \ref{sec:multiaperture}. Icon 4 clears all labels and apertures from the image display. Icon 5 starts the Multi-Plot module discussed in \S \ref{sec:multiplot}. Icon 6 opens previously saved photometry ``measurements tables'' (see Appendix \ref{app:measurementstable}). Icon 7 opens the \textit{Data Processor} panel discussed in \S \ref{sec:dataprocessor}. Icon 8 opens the \textit{Coordinate Converter} panel discussed in Appendix \ref{sec:coordinateconverter}.}
\label{fig:aijtoolbar}
\end{center}
\end{figure*}

When AIJ is started, the AIJ Toolbar opens and presents the eight AIJ-specific toolbar icons labeled 1 through 8 in Figure \ref{fig:aijtoolbar}. Those icons provide direct access to AIJ analysis tools or functions, including single aperture photometry (see \S \ref{sec:singleaperture}), aperture photometry settings (see Appendix \ref{sec:photset}), multi-aperture differential photometry (see \S \ref{sec:multiaperture}), multi-curve plotting (see \S \ref{sec:multiplot}), image calibration (see \S \ref{sec:dataprocessor}), and astronomical coordinate and time conversion (see Appendix \ref{sec:coordinateconverter}).

The 12 icons to the left of the AIJ icons and all of the menu options are standard IJ tools. These tools can also be used in AIJ, but normally only the \textit{File} menu options are needed for typical time-series data reductions. If all images in a sequence will not fit into the computer memory allocated to AIJ, the sequence can be opened as a ``virtual stack''. In this mode, the stack of images can be processed as if all images exist in memory, but AIJ loads only the single active/displayed image into memory. Virtual stacks perform more slowly than standard stacks, but memory requirements are minimal. All AIJ settings are persistent across sessions. Settings for specific configurations can be saved and reloaded later as needed. 

\subsection{Astronomical Image Display}\label{sec:aijdisplay}

\begin{figure*}
\begin{center}
\resizebox{0.85\textwidth}{!}{
\includegraphics*{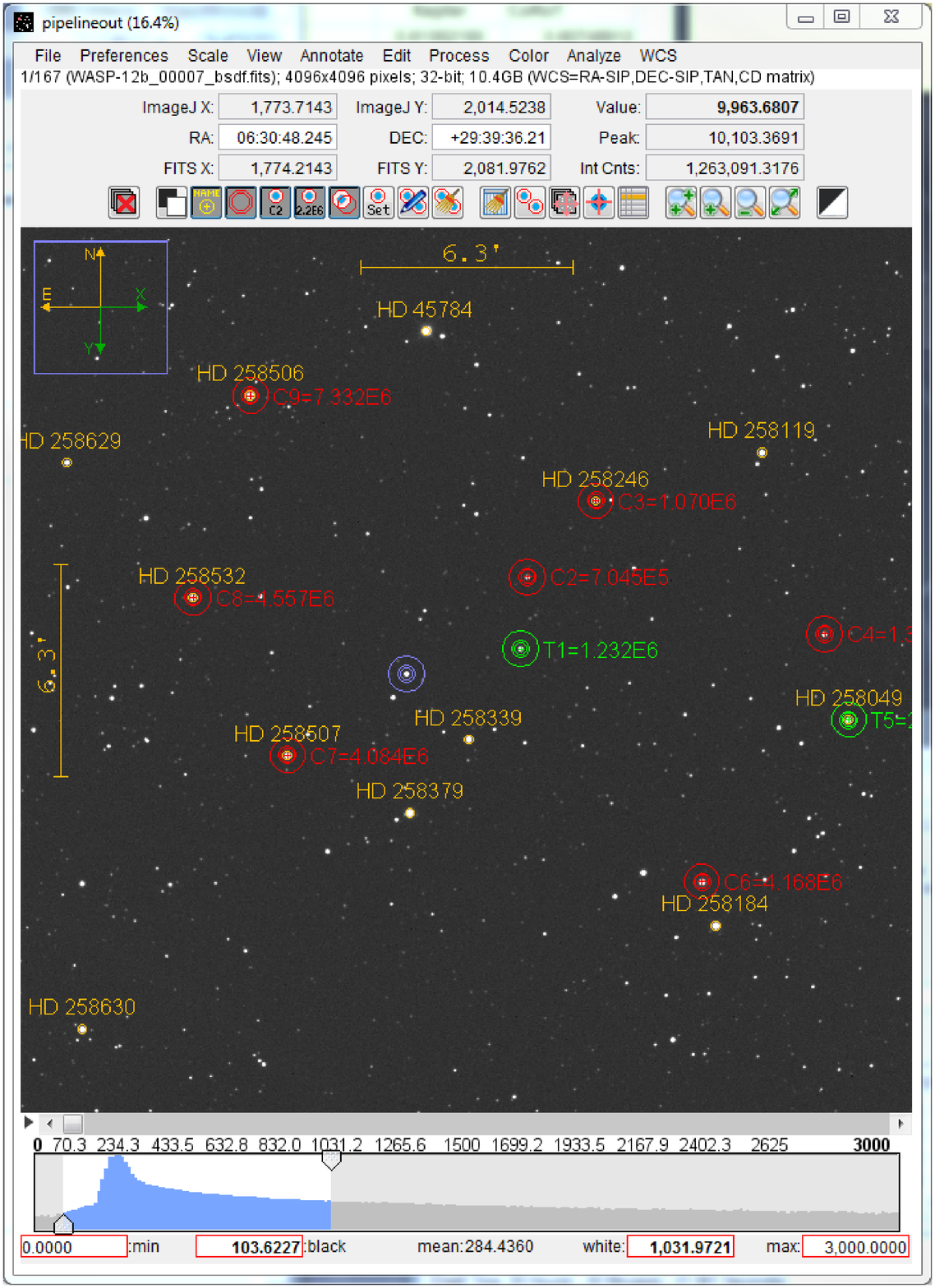} }
\caption{The AIJ image display. A wide range of astronomy specific image display options and image analysis tools are available from the menus, quick access icons, and interactive histogram. See text for details.}
\label{fig:aijimagedisplay}
\end{center}
\end{figure*}

Many popular image file formats are supported by AIJ, including the Flexible Image Transport System (FITS; \citealt{Wells:1981,Pence:2010}) file format. The astronomical image display shown in Figure \ref{fig:aijimagedisplay} offers numerous display options useful to astronomers. A menu system is available at the top of the image display window to provide access to all astronomy specific AIJ features. A row of quick access icons for control of frequently used image display options and image analysis tools is located directly above the image. Pixel and World Coordinate System (WCS; \citealt{Greisen:2002,Calabretta:2002,Greisen:2006}) information describing the image location pointed to by the mouse cursor is displayed in the three rows above the quick access icons. Image and WCS format information is displayed in the space under the image menus. A non-destructive image overlay optionally displays active apertures (green = target, red = comparison), object annotations, plate scale, and image orientation on the sky. 

The \textit{Scale} menu above an image display offers options for the control of image brightness and contrast (i.e. image scale). By default, image scale is set automatically and linearly maps the pixel values in the range $mean-0.5\sigma$ through $mean+2\sigma$ to 256 shades of gray running from black through white.

The zoom setting of an image display is most easily changed by rolling a mouse wheel. Image pan is controlled by a click and drag in the image, and a middle-click centers the clicked location in the image display. Arc-length can also be measured and displayed. An azimuthally averaged radial profile (i.e. a seeing profile) plot can be produced by an alt-left-click near an object (see Appendix \ref{sec:radialprofile}).

The \textit{View} menu above an image provides settings to invert the display of an image in \textit{x} and/or \textit{y}, enable or disable the display of the zoom indicator, the \textit{X}, \textit{Y}, \textit{N}, and \textit{E} directional arrows, and the image plate scale indicators in the image overlay. If WCS header information is available, AIJ automatically calculates and displays the $x$- and $y$-axis plate scales and the orientation of the image on the sky. Otherwise, those settings can be entered by the user. 

The blue aperture shown near the center of Figure \ref{fig:aijimagedisplay} moves with the mouse pointer. The value of the pixel at the mouse pointer and the peak pixel value and background-subtracted integrated counts (see \S \ref{sec:singleaperture}) within the mouse pointer aperture are updated in the right-hand column of data displayed above the image. This interactive mouse photometer helps to quickly assess which stars are suitable comparison stars during differential photometry set up. When AIJ is used in real-time data reduction mode (see \S \ref{sec:dataprocessor}), the mouse photometer helps to quickly determine an appropriate image exposure time and telescope defocus setting.

If a time-series of images is opened as an image stack, a scroll bar is available directly under the image display as shown in Figure \ref{fig:aijimagedisplay}, allowing the user to select which image of the stack is currently displayed. The right-pointing ``play'' icon to the left of the scroll bar will animate the image sequence at a predefined rate. 

An annotation feature allows objects to be labeled non-destructively in an image overlay. New object annotations can be added manually, or if the image has WCS information, target names can be extracted from SIMBAD and displayed by right-clicking on an object.

The quick access icons shown immediately above the image in Figure \ref{fig:aijimagedisplay} control which components of apertures are displayed (see \S \ref{sec:singleaperture}), control whether an aperture is to be centroided on a per aperture basis, provide direct access to the Multi-Aperture, Stack-Aligner, Astrometry, and FITS header editor modules, and control zoom and automatic contrast/brightness scaling.

\subsection{Utilities, Algorithms, and Measurements Tables}
In Appendix \ref{app:utilities}, we describe the following integrated utilities: Coordinate Converter (\ref{sec:coordinateconverter}), Astrometry/Plate Solving (\ref{sec:astrometry}), Image Alignment (\ref{sec:stackaligner}), Radial Profile plotting (\ref{sec:radialprofile}), Photometry Settings (\ref{sec:photset}), and Data Processor FITS Header Updates (\ref{sec:fitsheaderupdates}).

In Appendix \ref{app:photerr} we describe photometric uncertainty calculations. In Appendix \ref{app:apparentmagnitude} we describe the optional apparent magnitude and apparent magnitude uncertainty calculations. In Appendix \ref{app:measurementstable} we describe the measurements table used to store photometric results. 

\section{Data Processor: Image Calibration and Reduction}\label{sec:dataprocessor}

The Data Processor (DP) module provides tools to automate the build of master calibration images and the calibration of image sequences, and optionally perform differential photometry and light curve plotting. DP operates much like a script in that it processes selected calibration and science images in a user defined manner. Fields are provided to define the directory/folder locations and filename patterns of data to be processed (i.e. science, bias, dark, and flat images). The science image files can be further filtered based on the image sequence numbers contained in the filename. Controls are provided to enable various tasks that can be included in the data processing session. Disabling certain options will automatically disable other related input fields as appropriate to help the user understand which input fields are interconnected. File paths and names can be dragged and dropped from the OS into a field of the DP panel to minimize typing. 

Master bias, dark, and flat-field images can be created from raw images, or previously built master calibration files can be specified, for calibration of the science images. AIJ provides the option to either average or median combine the raw images when producing the master calibration files. Bias subtraction, dark subtraction, and flat-field division can be individually enabled. If bias subtraction is enabled, dark subtraction scaling can be enabled to scale the master dark pixel values by the ratio of the science image exposure time to the master dark image representative exposure time.

DP provides the option to implement CCD nonlinearity correction. This option replaces each pixel's ADU value in the bias-subtracted dark, flat, and science images with the corrected ADU value:
\begin{equation}
\rm{ADU_{corrected}}=c_0 + c_1\rm{ADU} + c_2\rm{ADU}^2  + c_3\rm{ADU}^3,
\label{eq:nonlinearity}
\end{equation}
\noindent where the coefficients $\rm{c_n}$ describe the non-linear behavior of the detector. Bias subtraction must be enabled to use non-linearity correction. Outlier pixel values can be removed using thresholded median filtering to compensate for artifacts in science images. This option is useful for improving the cosmetic appearance of images (e.g. to clean cosmic ray hits and/or hot and cold pixels). 

DP also provides options to calculate new astronomical data (e.g. airmass, time in \bjdtdb, target altitude, etc.) and add those data to the calibrated science image's FITS header (see Appendix \ref{sec:fitsheaderupdates}). DP can optionally run astrometry and add the resulting WCS data to the calibrated (and optionally to the raw) science image's FITS header.

Calibrated images can be output in a variety of file formats and image pixel depths (i.e. 16-bit integer or 32-bit floating point). Multi-Aperture (see \S \ref{sec:multiaperture}) and Multi-Plot (see \S \ref{sec:multiplot}) can be executed after each image is calibrated to perform differential photometry and display a light curve as the image data are processed. This feature is particularly useful for real-time reduction of data at the telescope. AIJ can also output the light curve plot and image display to a file after each science image is calibrated. These files can be used to update websites to show the progress of observations. A detailed log of all processing steps with timestamps is created by default.

\section{Photometry and Light Curve Capabilities}\label{sec:photometry}

AIJ provides interactive interfaces for single aperture photometry and multi-aperture differential photometry. The differential photometry interface is designed to automatically process a time-series of images and measure the light curves of exoplanet transits, eclipsing binaries, or other variable stars, optionally in real time as observations are being conducted.   

\subsection{Single Aperture Photometry}\label{sec:singleaperture}

Single aperture photometry measures the flux from a source within a predefined region of interest in an image referred to as an aperture. AIJ currently supports circular apertures only. A representation of an object's flux in the aperture, referred to as net integrated counts, is calculated by summing all of the pixel values within the aperture after subtracting an estimate of the background flux near the aperture. The background flux is estimated from the pixel values in a background annulus centered on the aperture. Single aperture photometry can be performed by simply placing the mouse pointer near the center of an object in an image. The net integrated counts within the mouse pointer aperture are shown in the display area above the image. The results of a single aperture measurement can also be recorded in a ``measurements table'' (see Appendix \ref{app:measurementstable}).

AIJ calculates photometric uncertainty as described in Appendix \ref{app:photerr}. For proper uncertainty calculations, the gain, dark current, and read out noise of the CCD detector used to collect the data must be entered before conducting photometric measurements.

\subsection{Multi-Aperture Differential Photometry}\label{sec:multiaperture}

Differential photometry measures the flux of a target star relative to the combined flux of one or more comparison stars. The differential measurement is conducted by performing single aperture photometry on one or more target stars and one or more comparison stars. Then a target star's differential flux is calculated by dividing the target star's net integrated counts, $F_{T}$, by the sum of the net integrated counts of all comparison stars (i.e. the sum of $F_{C_i}$, where $i$ ranges from 1 to the number of comparison stars $n$). The calculation is:
\begin{equation}
\rm{rel\_flux\_T\_j} =  \frac{F_T}{\sum_{i=1}^{n} F_{C_i}},
\label{eq:difffluxT}
\end{equation}
\noindent where $j$ indicates the target star aperture number and $i$ indexes all comparison star aperture numbers. The terms relative flux and differential flux are used in this work and in AIJ interchangeably. Differential photometric error is calculated as described in Appendix \ref{app:photerr}.

AIJ also calculates differential flux for each comparison star aperture by comparing the flux in its aperture to the sum of the flux in all \textit{other} comparison star apertures. The calculation is:
\begin{equation}
\rm{rel\_flux\_C\_j} =  \frac{F_{C_j}}{\sum_{i=1}^{n} F_{C_i}, i\ne j},
\label{eq:difffluxC}
\end{equation}
\noindent where $j$ indicates the comparison star aperture number for which differential flux is being calculated and $i$ indexes all comparison star aperture numbers. 

The Multi-Aperture (MA) module automates the task of performing differential photometry on a time-series of images. Various settings are presented in a set-up panel, and then the target and comparison star apertures are placed and adjusted interactively by clicking near stars directly in the image display. The differential photometry results are recorded in a ``measurements table'' (see Appendix \ref{app:measurementstable}). If WCS headers are available, apertures can optionally be placed first according to RA and Dec coordinates, rather than by $x$ and $y$ pixel coordinates, and then optionally centered on the nearest star. Aperture definitions can be stored and reopened from disk.

AIJ supports both fixed and variable radius apertures. For the fixed radius option, the user specifies the aperture radius and sky-background inner and outer radii to be used in all images of the time series.  For the variable radius option, the aperture radius used for a specific image of a time-series is calculated using one of two methods. The first method determines the per image radius as the product of a user specified full-width at half-maximum (FWHM) factor and the average FWHM from all apertures in that image. The second method determines the aperture radius from an azimuthally averaged radial profile (see Appendix \ref{sec:radialprofile}), centered on the aperture. In this mode, the aperture radius used in an image is equal to the distance from the center of the aperture at which the radial profile value is equal to a user specified normalized flux cutoff. The variable apertures may improve photometric precision when seeing varies significantly or telescope focus drifts within a time-series. The variable aperture modes should not be used in crowded fields since the changing aperture radius will worsen the effects of variable amounts of contaminating flux blending into the aperture as seeing or focus changes. 

By default, iterative $2\sigma$ cleaning of the sky-background region is performed to identify and reject pixels containing stars or other anomalies. The iteration continues until the mean ADU of the pixels remaining in the background set converges or the maximum number of allowed iterations has been reached without convergence. The sky-background pixels remaining after the cleaning operation are used to calculate and remove the sky-background at each pixel in the aperture. AIJ can optionally fit a plane to the remaining pixels in the background region and subtract the value of the plane at each pixel within the aperture to remove the sky-background contribution. Otherwise, AIJ subtracts the mean of the remaining pixels in the background annulus from each pixel in the  aperture.

Apparent magnitudes of target aperture sources can optionally be calculated from user entered apparent magnitudes of one or more comparison aperture sources (see Appendix \ref{app:apparentmagnitude}). AIJ provides an option to output the apparent magnitude data in a format compatible with the Minor Planet Center. 

\subsection{Multi-Plot}\label{sec:multiplot}

The Multi-Plot (MP) module provides a multi-curve plotting facility that is tightly integrated with differential photometry and light curve fitting. MA can automatically start MP, or MP can be manually started by clicking the associated icon on the AIJ Toolbar. If a measurements table has been created by MA or opened from the OS, MP will automatically create a plot based on the last plot settings. Alternatively, plot templates can be saved and restored to easily format commonly created plots. Plotting controls are accessed in two user interface panels.

The main plotting panel contains settings that affect the overall plot, including the plot title and subtitle, legend placement and options, $x-$ and $y-$axis label and scaling options, and the plot size in pixels. The main panel also provides access to other $x$-axis controls that define the regions used when normalizing, detrending, and fitting $y$-axis datasets. If applicable, the time of telescope meridian flip can be specified to allow detrending of any baseline offset from one side of the meridian flip to the other. As discussed in Appendix \ref{sec:coordinateconverter}, new astronomical data such as AIRMASS, $\bjdtdb$, etc. can be calculated and added to the measurements table using the main panel. 

\begin{figure*}
\begin{center}
\resizebox{0.85\textwidth}{!}{
\includegraphics*{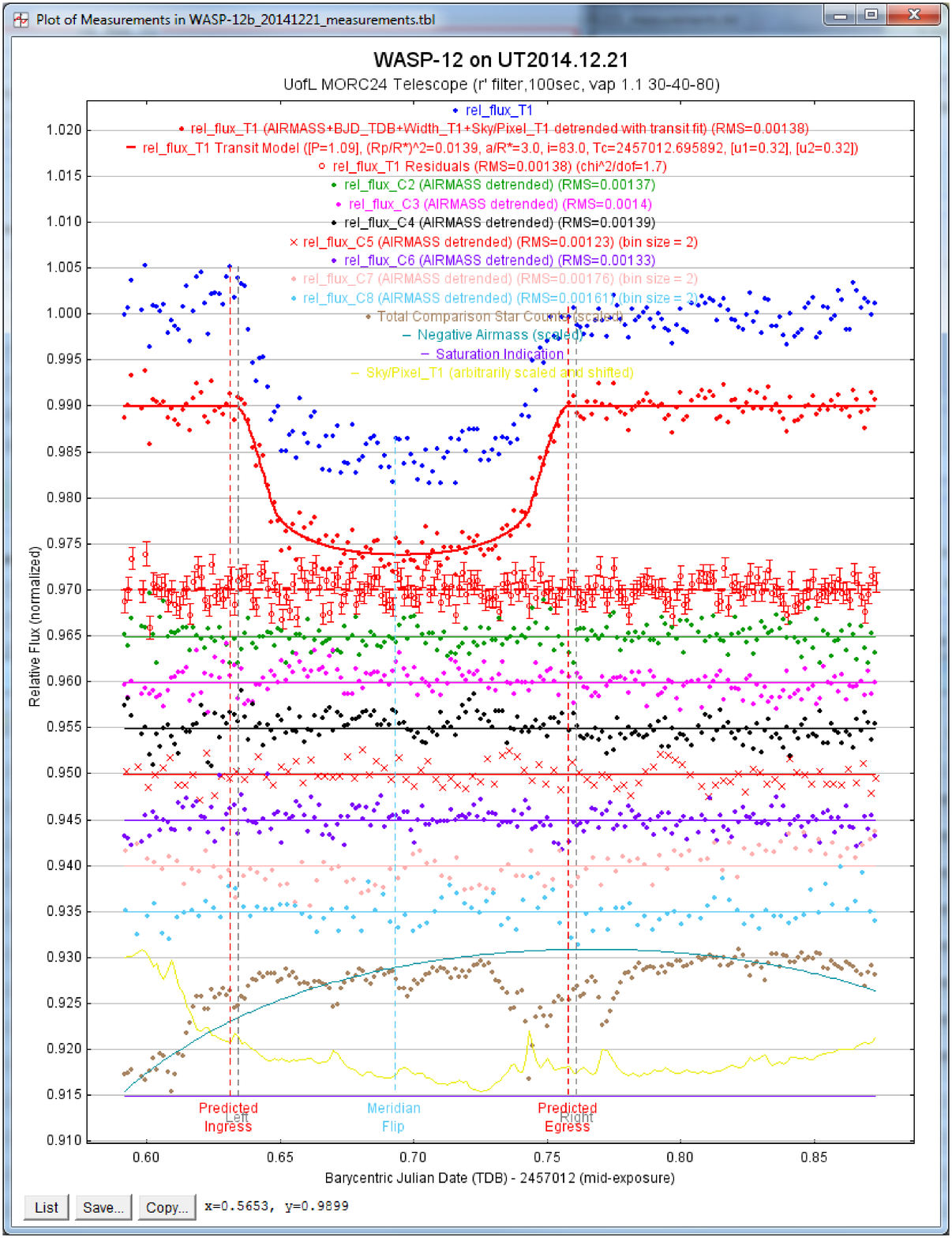} }
\caption{Multi-Plot example plot. The plot of a WASP-12b transit under poor observing conditions is shown. The top dataset plotted with solid blue dots is the raw normalized differential photometry. The solid red dots show the light curve after simultaneously detrending and fitting with an exoplanet transit model, which is shown by the red line through the data. Note the reduced systematics and scatter in the detrended data. The open red dots show the model residuals with error bars. See the text for descriptions of the other plotted data.}
\label{fig:aijmpplot}
\end{center}
\end{figure*}

In a secondary plotting panel, multiple rows of settings are available to control how datasets are plotted. Each row of settings controls an individual plotted dataset. The number of rows available (i.e. the maximum number of plotted datasets) is user configurable. Each row of plot settings controls selection of the $x$- and $y$-axis datasets plotted, color and symbol used, the number of data points to average combine, the curve fitting mode, the detrend datasets, the normalization (or magnitude reference) region mode, scaling and shifting values, and legend settings. 

Figure \ref{fig:aijmpplot} shows an example plot of a WASP-12b transit and demonstrates many of MP's plotting capabilities. The raw normalized light curve is shown as solid blue dots near the top of the plot. In this case, the in-transit data are excluded from the calculation of the normalization parameter. The solid red dots show the same light curve after simultaneously detrending and fitting the data to an exoplanet transit model (see \S \ref{sec:fitanddetrend}) and shifting down by 0.010 on the $y$-axis.  Note the reduced systematics and scatter in the data. The light curve model residuals are shown as open red circles with AIJ calculated error bars.

The next seven datasets plotted below the residuals (green, magenta, and black solid dots, red $\times$-symbol, and pink and light blue filled dots) show the first seven comparison star differential light curves, which have all been airmass detrended assuming a constant brightness model. Note that comparison stars 5, 7, and 8 have been binned by 2 data samples to reduce the scatter for plotting purposes. The current implementation of binning is actually averaging of the specified number of data points, rather than binning into fixed size $x$-axis bins.

The four datasets over-plotted at the bottom of the figure (cyan, yellow, and purple lines, and brown dots) show four diagnostic curves (see legend at top of Figure \ref{fig:aijmpplot}). These curves are plotted using a mode that scales the minimum and maximum values of the dataset to fit within a specified percentage of the vertical size of the plot after shifting by a specified percentage of the vertical size of the plot. This mode makes scaling of data to fit on a plot easy when the shape of a curve is important, but the actual values of the data are not.

Datasets displayed in a plot have typically been modified in one or more ways (e.g. normalized, detrended, converted to/from magnitude, scaled, shifted, binned, etc.). Those values are calculated on-the-fly when a change is made to a plot setting, but do not appear in the measurements table. However, MP provides an option to allow the user to add the modified/displayed values to the measurements table as new data columns for further manipulation or permanent storage. Model residuals and sampled versions of the model can also be saved to the measurements table.

The legend is shown at the top of the plot in Figure \ref{fig:aijmpplot}. Legend entries for each plotted curve can be automatically generated based on the measurements table data column names known to be produced by MA, or a custom legend can be displayed individually or in combination with the automatic legend. By default, the RMS of the model residuals is calculated and displayed in the legend for fitted and/or detrended light curves. The light curve model parameter values are optionally shown for fitted light curves. The predicted time of ingress and egress are shown as red dashed vertical lines. The meridian flip time is indicated by the light blue vertical dashed line (although no meridian flip actually occurred for the example observations), and the \textit{Left} and \textit{Right} gray dashed vertical lines show the boundaries of the normalization, detrending, and fitting regions.

All typical data and image products created by a photometry and/or plotting and fitting session can be saved with one action using the \textit{Save All} feature in the MP main panel. There are also \textit{File} menu options in the main panel and image display panels to save each data product separately. 

\subsection{Light Curve Fitting and Detrending}\label{sec:fitanddetrend}

\begin{figure*}
\begin{center}
\resizebox{\textwidth}{!}{
\includegraphics*{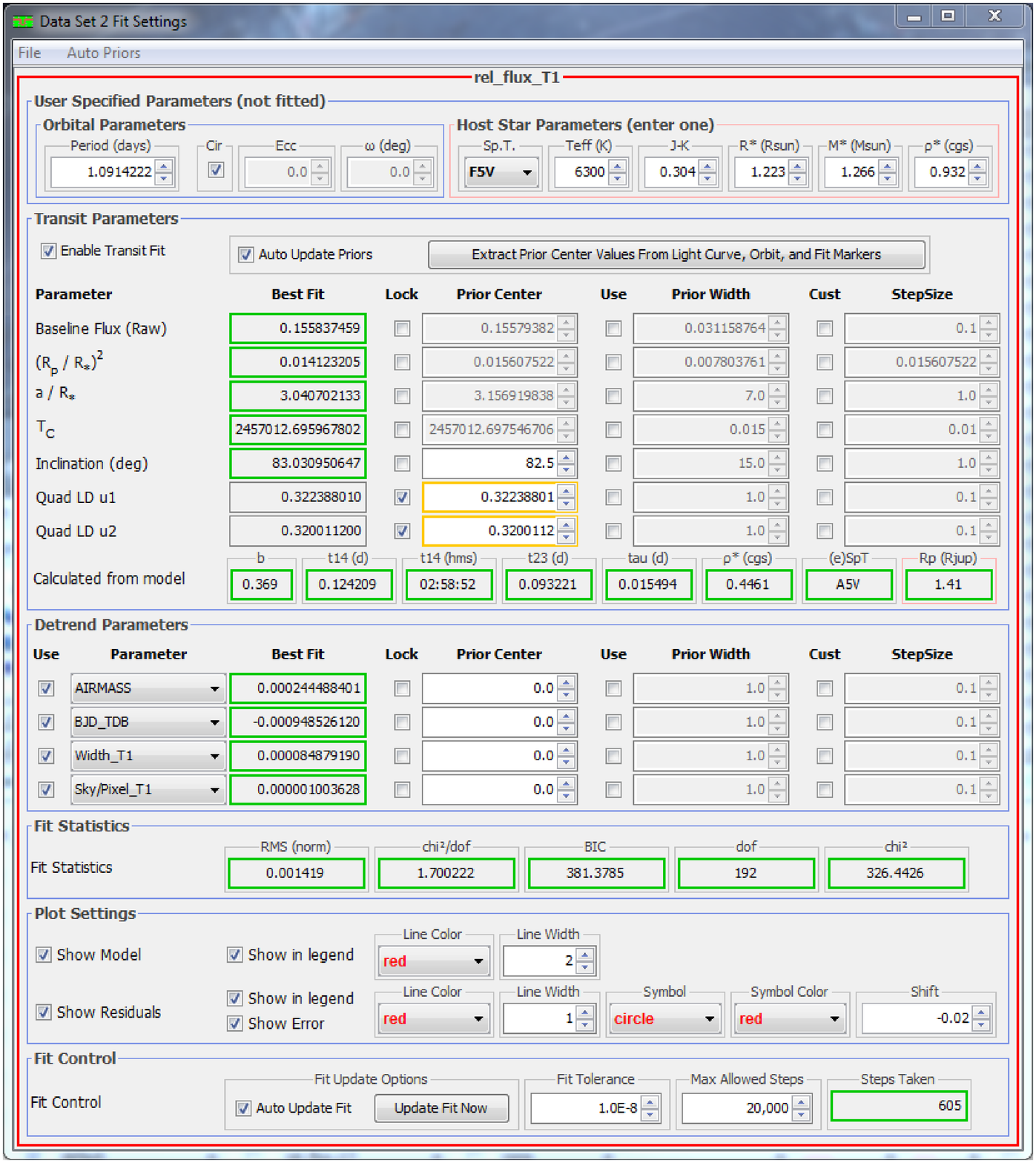} }
\caption{The \textit{Fit Settings} panel. The light curve model fitting and detrending panel settings shown here produce the light curve model fit in the example plot of Figure \ref{fig:aijmpplot}. Light curve prior center and width values can optionally be set by the user to constrain the model fit. Direct access to all detrending parameters is also provided along with more flexible settings. Several values calculated from the model parameters are displayed, along with several statistical values useful for assessing data quality and goodness of the model fit to the data. See text for more details.}
\label{fig:aijfitpanel}
\end{center}
\end{figure*}

Light curve fitting is enabled for a particular dataset in the MP secondary panel by selecting the transit fit mode in the row of plotting controls for that dataset. When this mode is selected, a \textit{Fit Settings} panel will be displayed for the dataset as shown in Figure \ref{fig:aijfitpanel}. The settings in the figure produce the light curve model fit shown in the example plot of Figure \ref{fig:aijmpplot}. The transiting exoplanet model is described in \citet{Mandel:2002}. The transit is modeled as an eclipse of a spherical star by an opaque planetary sphere. The model is parametrized by six physical values, plus a baseline flux level, $F_0$. The six physical parameters are the planetary radius in units of the stellar radius, $R_P/R_*$, the semi-major axis of the planetary orbit in units of the stellar radius, $a/R_*$, the transit center time, $T_C$, the impact parameter of the transit, $b$, and the quadratic limb darkening parameters, $u_1$ and $u_2$. The orbital inclination can be calculated from the model parameters as
\begin{equation}
i=\cos^{-1}\left(b\frac{R_*}{a}\right).
\end{equation} 

AIJ is currently limited to finding the best fit model parameter values and does not provide estimates of the parameter uncertainties. The best fit model is found by minimizing $\chisq$ of the model residuals using the downhill simplex method to find local minima \citep{Nelder:1965}.

The parameter settings in the \textit{User Specified Parameters} sub-panel are not fitted. The period of the exoplanet's orbit is not well constrained by the transit model, but its value will impact the best fit values of some of the fitted parameters, so the period must be manually entered by the user. Normally, the orbital period is known from RV or wide-field survey photometric data. The \textit{Host Star Parameters} in the same sub-panel are only used to estimate the physical planet radius, $R_P$, from the fitted parameter $R_P/R_*$. The host star parameter values are interrelated by tables in \citet{Allen:2001} for zero age main sequence (ZAMS) stars. The only value used in the calculation of $R_P$ (displayed near middle of the right-hand side of the panel) is $R_*$, so that value should be entered directly if known. Otherwise, entering any one of the other host star parameters will produce a rough estimate of $R_*$ based on the ZAMS assumption.

The \textit{Transit Parameters} sub-panel has seven rows for the seven transit model parameters. \textit{Prior Center} values will need to be set for the seven parameters to ensure the correct $\chisq$ minimum is found. The top four parameters shown in the sub-panel are extracted from the light curve data by default. In the odd case that those estimated values do not produce a proper fit, the values can be entered directly by the user. In the example shown, the \textit{Inclination} prior center value has been set by the user, but no constraints have been placed on the range of valid final fitted values (although the upper end is limited to $90\degrees$ by the definition of inclination). The \textit{Quad LD u1} and \textit{Quad LD u2} parameter values have been set by the user, and the fitted values have been locked to those values by enabling the \textit{Lock} option beside each one. The fixed values of u1 and u2 were extracted from the \citet{Claret:2011} theoretical models using a website tool\footnote{http://astroutils.astronomy.ohio-state.edu/exofast/limbdark.shtml}. The best fit transit model parameter values are displayed in the \textit{Best Fit} column. A green box around the fitted parameter values indicates that the minimization converged to a value less than the \textit{Fit Tolerance} within the \textit{Max Allowed Steps}. Both of those minimization parameters can be set at the bottom of the fit panel in the \textit{Fit Control} sub-panel, but the default values normally work well.

The bottom row in the \textit{Transit Parameters} sub-panel shows several values that are calculated from the best fit model. The host star's density, $\rho_*$, is particularly interesting, since a good estimate can be derived from the transit light curve data alone. Positioning the mouse pointer over any parameter will optionally cause a description of the parameter to be temporarily displayed.

The \textit{Prior Width} column allows the user to optionally limit the range of a parameter's fitted value. \textit{Prior Width} values are not normally needed, but may be helpful in fitting an ingress- or egress-only partial transit. The \textit{StepSize} column allows the user to set a custom initial minimization step size. However, the default values for each parameter normally work well, so setting custom values is not usually necessary.

The \textit{Detrend Parameters} sub-panel duplicates the detrend settings in the MP secondary panel. However, the \textit{Fit Settings} panel provides direct access to all detrend parameters and settings. Prior center values, widths, and fitting step sizes can optionally be set for detrend parameters as well. 

Light curve detrending is accomplished by including a $\chisq$ contribution for each selected detrend parameter in the overall light curve fit. The $\chisq$ contribution at each step of the minimization represents the goodness of the linear fit of the detrend parameters to the light curve after subtracting the light curve model corresponding to the current fit step. The $\chisq$ contribution for all $n$ detrend parameters is calculated at each step of the fitting process as
\begin{equation}
\chisq_{D} = \sum_{k=1}^{m} \frac{\left(O_k-\left(\sum_{j=1}^{n}c_{j}D_{j_k}\right)-E_k\right)^2}{\sigma_k^2},
\label{eq:chi2fordetrend}
\end{equation}
\noindent where $j$ indexes the detrend parameters, $k$ indexes the samples of the light curve, $m$ is the total number of samples in the light curve, $O_k$ is the observed normalized differential target flux, $c_j$ is the fitted linear coefficient for the detrend parameter values $D_{j_k}$, $E_k$ is the expected value of the flux (which is the normalized transit model value corresponding to the time of the $k^{\rm th}$ data sample), and $\sigma_k$ is the error in the normalized differential target flux for each sample.

The \textit{Fit Statistics} sub-panel lists five statistical values that allow the user to assess the quality of the data and goodness of the model fit to the data. The values displayed from left to right are RMS of the model residuals, $\chisq$ per degree of freedom (i.e. reduced $\chisq$), Bayesian Information Criterion (BIC), the number of degrees of freedom, and the total $\chisq$. BIC is defined as
\begin{equation}
{\rm{BIC}}=\chisq+p\ln n,
\end{equation}
\noindent where $p$ is the number of fitted parameters, and $n$ is the number of fitted data points. The BIC can be used to determine whether the addition of a new parameter to a model (in particular an optional one such as a detrend parameter) provides a significant improvement in the fit. If the BIC value decreases by more than 2.0 when a model parameter is added, then the new model is preferred over the model with fewer parameters. A larger decrease in the BIC value suggests a stronger preference for the new model.

\subsection{Comparison Ensemble Management}\label{sec:ensemblemanagement}

The MP environment allows the user to include or exclude comparison stars from the comparison ensemble without re-running Multi-Aperture, as long as apertures were defined for all potentially good comparison stars in the original Multi-Aperture run. Excluded stars are considered target stars. When a star is added to or removed from the ensemble, the relative flux values for each star are recalculated and the measurements table, model fit, and plot are updated. 

An option allows the user to quickly cycle through the comparison ensemble removing one star at a time so that poor comparison stars can be quickly identified and removed from the ensemble. Another option enables a single comparison star and cycles through each one to allow the user to quickly assess the quality of each comparison star individually.

\acknowledgments
K.A.C. and J.F.K acknowledge support from NASA Kentucky Space Grant Consortium and its Graduate Fellowship program. 
K.A.C. and K.G.S. acknowledge support from NSF PAARE grant AST-1358862 and the Vanderbilt Initiative in Data-intensive Astrophysics.  
We would like to acknowledge the support of early users of AIJ, who's feedback helped drive the development of the current feature set and resolve software bugs and reliability issues.
We thank the anonymous referee for a thoughtful reading of the manuscript and for useful suggestions. 
This work has made use of the SIMBAD database operated at CDS, Strasbourg, France.

\appendix

\section{AIJ Utilities}\label{app:utilities}

\subsection{Coordinate Converter}\label{sec:coordinateconverter}

The Coordinate Converter (CC) module converts astronomical coordinates and times to other formats based on observatory location, target coordinates, and time of observation. CC can be operated as a module under full control of the user, and it can be operated under the control of DP and MP to provide automated calculations within those modules. Detailed CC help is available in the menus above the CC panel.

The target coordinates can be specified by entering a Simbad resolvable object name, or by entering J2000, B1950, or time of observation equatorial RA and Dec, J2000 or time of observation ecliptic longitude and latitude, galactic longitude and latitude, or horizontal altitude and azimuth. Any input coordinate format is converted to all other coordinate formats, plus hour angle, zenith distance, and airmass.

The time of observation can be specified by entering UTC time and date, local time and date, JD (UTC-based), HJD (UTC-based), or BJD (TDB-based). Time entered in any format is converted to all other formats, plus local sidereal time (LST). Also, AM and PM nautical twilight times are displayed for the specified day. The current time is always displayed as UTC, local, JD (UTC-based), and LST.

The $\bjdtdb$ time format requires dynamical time as the time-base. Dynamical time accounts for the changing rotational speed of the Earth by implementing leap-seconds. Leap-second updates are not periodic, but are announced six months before taking effect. New leap-seconds can be added to AIJ's leap-second table from The U.S. Naval Observatory website.

DP creates an instance of CC (DPCC), and depending on user settings, data can be extracted from FITS header information or entered manually to control the settings used by DPCC to calculate new astronomical values to add to the calibrated image's FITS header information (see Appendix \ref{sec:fitsheaderupdates}). If the FITS header contains the time of observations, the target's SIMBAD ID or coordinates, and the observatory's ID or coordinates, DPCC calculations can be executed with no user input. If target and/or observatory information is not available in the header, the missing information can be entered by the user.

MP also creates an instance of CC (MPCC) to allow new astronomical values to be added to a measurements table. Time and object coordinate values can be extracted from table columns and used to calculate any of the supported time and coordinate formats mentioned above.

The phase of the moon, and the altitude and proximity of an object to the moon and solar system planets is also displayed along the bottom of the CC panel to help with observation planning, or to illustrate those conditions during a time series.

\subsection{Astrometry/Plate Solving}\label{sec:astrometry}

The astrometry feature ``plate solves'' images using an internet connection to the astrometry.net web portal at \url{nova.astrometry.net} \citep{Lang:2010}. AIJ searches the image and extracts the source locations. Only the $x,y$ coordinates for a subset of the brightest extracted sources are sent to \url{nova.astrometry.net}. The actual image is not transferred across the network, which limits network traffic and improves the solve time. After a successful astrometric solution is found, WCS headers are automatically added to the FITS image header, and the file can optionally be resaved with the WCS solution.

Images can be blindly solved with no knowledge of the sky coordinates or plate scale of the image. However, solve time may be faster if the plate scale is known and entered by the user. If the approximate sky coordinates of the center of the image are known, entering those values may also improve solve time. By default, a log file is created to record the results of the plate solve for each image in the stack. When a field is successfully solved, astrometry.net returns a list of sources that are in the image. The source names and locations can be displayed in the image and/or saved to the FITS header. The plate solve process takes $\sim10-20$ seconds per image.

\subsection{Image Alignment}\label{sec:stackaligner}

The images within a stack can be aligned using the Stack Aligner module. At the time of writing, Stack Aligner only supports image translation for alignment. Image rotation and scaling are not currently implemented. 

Stack Aligner provides two methods to align images. If all images in the stack have been plate solved, the images can be aligned using information in the WCS headers. If images have not been plate solved, apertures may be used to identify alignment stars. Aperture placement is performed in the same way as described for Multi-Aperture in \S \ref{sec:multiaperture}, and images are aligned based on the average of alignment star centroid offsets between consecutive images. 

Another image alignment option uses an image stabilizer algorithm that is useful for alignment of non-stellar objects.  This tool will remove atmospheric jitter from a rapid sequence of planetary or lucky images, or track a comet over a long duration as it moves across a star field.

\subsection{Radial Profile}\label{sec:radialprofile}

\begin{figure}
\begin{center}
\resizebox{\columnwidth}{!}{
\includegraphics*{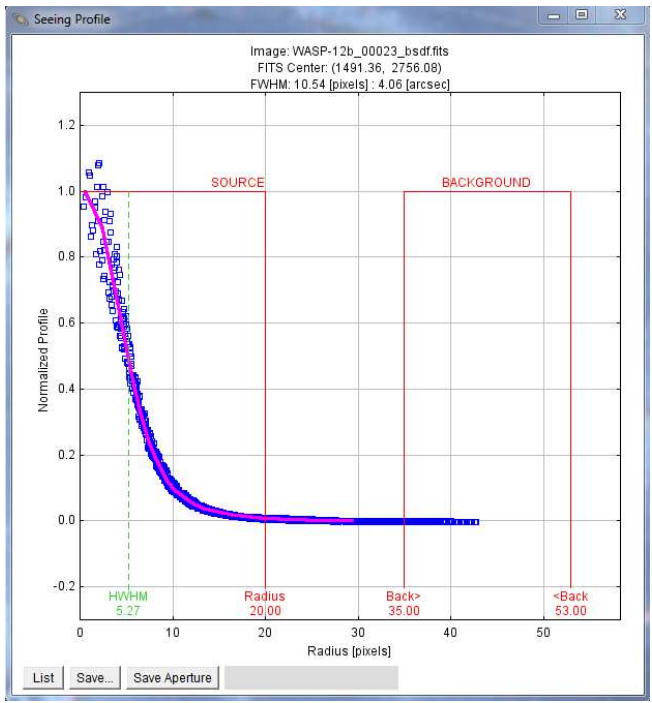} }
\caption{Radial Profile (Seeing Profile) plot. The plot shows the half-width at half-maximum (HWHM), the FWHM, and suggested aperture radii in pixels. The aperture radius is set to $1.7\times$FWHM, the inner radius of the sky-Abackground annulus is set to $1.9\times$FWHM, and the outer radius of the sky-background annulus is set to $2.55\times$FWHM. The FWHM is also given in seconds of arc if valid WCS headers are available.}
\label{fig:aijseeingprofile}
\end{center}
\end{figure}

AIJ can produce a plot of the azimuthally averaged radial profile of an object in an image.  Figure \ref{fig:aijseeingprofile} shows an example radial profile plot. The plot shows the half-width at half-maximum (HWHM), the FWHM, and suggested aperture radii in pixels. The aperture radius is set to $1.7\times$FWHM, the inner radius of the sky-background annulus is set to $1.9\times$FWHM, and the outer radius of the sky-background annulus is set to $2.55\times$FWHM. These radii give an equal number of pixels in the aperture and sky-background annulus. The FWHM is also given in seconds of arc if valid WCS headers are available. The calculated aperture settings can be transferred to the photometer. 

\subsection{Photometry Settings}\label{sec:photset}

Settings related to photometric measurements are specified in two \textit{Aperture Photometry Settings} panels. The main panel provides access to the aperture radii, centroid, and background settings. A list of FITS keywords can be entered to specify that the corresponding numeric data in the image headers be extracted and added to the measurements table as part of a photometry run. The CCD gain, readout noise, and dark current must be entered in this panel for proper photometric error calculations. Linearity and saturation warning levels should be entered so that indicators of those conditions can we properly noted in the measurements table and AIJ's user interfaces. 

Two centroid methods are available. One method uses the algorithm in \citep{Howell:2006} and gives highly repeatable \textit{x,y} centroid results (i.e. is not sensitive to the \textit{x,y} starting location). The other option uses a center-of-mass (i.e. center-of-flux) algorithm and provides better results when placing apertures around defocused stars.   

The second panel allows the selection of photometric data items to be included in the measurements table. It is highly recommended to enable all data items since some AIJ functionality requires certain data items exist in the table. The maximum number of apertures allowed per image is set in this panel (500 by default).

\subsection{Data Processor FITS Header Updates}\label{sec:fitsheaderupdates}

DP provides the option to calculate new astronomical data and add it to the FITS header of a calibrated image. FITS header keywords can be specified for the extraction of header data from the raw images as input for the calculation of various new astronomical data values. The input keywords that can be specified are the object coordinates and the observatory latitude and longitude. The new values that can be calculated and added to a calibrated image header are the same as the target coordinate and observation time formats described in Appendix \ref{sec:coordinateconverter}. 

\section{Photometric Error Calculation}\label{app:photerr}

Proper estimation of the uncertainty (i.e. error or noise) in each photometric measurement is important for reporting the significance of the measurement and plotting error bars in the light curve plot, but it is also important for the proper calculation of the best fit model to the data, since the uncertainty of each measurement, $\sigma$,  is part of the $\chisq$ calculation used in the fitting process (e.g. see equation \ref{eq:chi2fordetrend}). In short, the $\chisq$ contribution from each data point is weighted by a factor of $1/\sigma^2$, which places more weight on data with small errors, and less weight on data with large errors.

\citet{Mortara:1981} and \citet{Howell:1989} discuss the noise contributions to the measurement of a point source using CCD aperture photometry and develop the ``CCD equation'' to estimate the signal-to-noise ratio of a measurement. \citet{Merline:1995} construct a computer model of the same measurement and develop the more rigorous ``revised CCD equation''. The equation gives the total noise N in ADU for a CCD aperture photometry measurement as:
\begin{equation}
N=\frac{\sqrt{GF_*+n_{pix}(1+\frac{n_{pix}}{n_b})(GF_S+F_D+F^2_R+G^2\sigma^2_f)}}{G},
\label{eq:ccdnoise}
\end{equation}
\noindent where $G$ is the gain of the CCD in electrons/ADU, $F_*$ is the net (background subtracted) integrated counts in the aperture in ADU, $n_{pix}$ is the number of pixels in the aperture, $n_b$ is the number of pixels in the region used to estimate sky background, $F_S$ is the number of sky background counts per pixel in ADU, $F_D$ is the total dark counts per pixel in electrons, $F_R$ is read noise in electrons/pixel/read, and $\sigma_f$ is the standard deviation of the fractional count lost to digitization in a single pixel ($\sigma_f\simeq0.289$ ADU for $f$ uniformly distributed between $-0.5$ and $0.5$).

The AIJ photometer automatically performs the noise calculation described by equation \ref{eq:ccdnoise} for each aperture. For proper noise calculation, the gain, dark current, and read out noise of the CCD detector used to collect the data must be entered in the \textit{Aperture Photometry Settings} panel (see Appendix \ref{sec:photset}). For differential photometry, AIJ propagates the noise from all apertures to derive the error in differential flux measurements. First, the noise from the apertures of each comparison star are combined in quadrature to give the total comparison ensemble noise:
\begin{equation}
N_E=\sqrt{\sum_{i=1}^{n}N_{C_i}^2},
\label{eq:ensemblenoise}
\end{equation}
\noindent where $i$ indexes the comparison stars of the ensemble, and $N_{C_i}$ is the noise for each comparison star as calculated by equation \ref{eq:ccdnoise}, and $n$ is the number of comparison stars. Error is then propagated through the relative flux quotient to find the relative flux error for the target star as:
\begin{equation}
\sigma_{\rm{rel\_ flux}}=\frac{F_T}{F_E}\sqrt{\frac{N_T^2}{F_T^2}+\frac{N_E^2}{F_{E}^2}},
\label{eq:relfluxerr}
\end{equation}
\noindent where $F_T$ is the net integrated counts in the target aperture, $F_E$ is the sum of the net integrated counts in the ensemble of comparison star apertures, $N_T$ is the noise in the target star aperture from equation \ref{eq:ccdnoise}, and $N_E$ is the ensemble noise from equation \ref{eq:ensemblenoise}.
\\
\section{Apparent Magnitude Calculation}\label{app:apparentmagnitude}

The apparent magnitude of target aperture sources can be calculated by entering the apparent magnitude of one or more comparison aperture sources. By default, if WCS information is available in the FITS image header, SIMBAD information is presented to assist in determining the comparison star apparent magnitude. 

The apparent magnitude of the source in target aperture \textit{nn} is calculated as
\begin{multline}\label{eq:targetappmag}
\mathrm{Source\_AMag\_T}nn =\\
\frac{-\ln \sum_{xx} 2.512^{-\mathrm{Source\_AMag\_C}xx}}{\ln2.512}~-\\
2.5 \log{\frac{\mathrm{Source\hbox{-}Sky\_T}nn}{\sum_{xx} \mathrm{Source\hbox{-}Sky\_C}xx}}, 
\end{multline}
\noindent where \textit{nn} is the target aperture number, \textit{xx} indexes all comparison apertures for which an apparent magnitude has been entered by the user, Source\_AMag\_C\textit{xx} are the user entered comparison source apparent magnitudes, and Source-Sky\_T\textit{nn} and Source-Sky\_C\textit{xx} are the net integrated counts in apertures \textit{nn} and \textit{xx} as defined in Appendix \ref{app:measurementstable} and \S \ref{sec:photometry}. 

The uncertainty in the apparent magnitude of the source in target aperture \textit{nn} is calculated as
\begin{multline}\label{eq:targetappmagerr}
\mathrm{Source\_Amag\_Err\_T}nn = 2.5 \log\left(1 + \right.\\
\left.\sqrt{\frac{\mathrm{Source\_Error\_T}nn^{\,2}}{\mathrm{Source\hbox{-}Sky\_T}nn^{\,2}} + \frac{\sum_{xx}\mathrm{Source\_Error\_C}xx^{\,2}}{\left(\sum_{xx}\mathrm{Source\hbox{-}Sky\_C}xx\right)^2}}\right), 
\end{multline} 
\noindent where \textit{nn} is the target aperture number, \textit{xx} indexes all comparison apertures for which an apparent magnitude has been entered by the user, Source-Sky\_T\textit{nn} and Source-Sky\_C\textit{xx} are the net integrated counts and Source\_Error\_T\textit{nn} and Source\_Error\_C\textit{xx} are the  net integrated counts uncertainties for apertures \textit{nn} and \textit{xx}. The target aperture apparent magnitude uncertainties do not include the uncertainty in the comparison source apparent magnitudes entered by the user.

The uncertainty in the apparent magnitude of the source in comparison aperture \textit{nn} is calculated as
\begin{multline}\label{eq:compappmagerr}
\mathrm{Source\_Amag\_Err\_C}nn =\\
2.5 \log\left(1 + \frac{\mathrm{Source\_Error\_C}nn}{\mathrm{Source\hbox{-}Sky\_C}nn}\right), 
\end{multline} 
\noindent where \textit{nn} is the comparison aperture number and Source-Sky\_C\textit{nn} and Source\_Error\_C\textit{nn} are the net integrated counts and the net integrated counts uncertainty, respectively, for comparison aperture \textit{nn}. The comparison aperture apparent magnitude uncertainty is simply the flux-based photometric error converted to the magnitude scale. These values do not include the uncertainty in the comparison source apparent magnitudes entered by the user. 

\section{Measurements Tables}\label{app:measurementstable}

The results from single aperture photometry and multi-aperture differential photometry are stored in a ``measurements table". The table can be exported as a tab or comma delimited text file input to other programs. For single aperture photometry, each row in a table contains all measurements produced from a single aperture. For multi-aperture differential photometry, a row contains all measurements produced by all apertures in a single image, and a row exists for each image that has been processed. Each column contains the same measurement from all images. Columns are labeled with unique names, and those names are available for selection in the pull-down menus in MP.

\bibliographystyle{apj}

\bibliography{AstroImageJ}

\end{document}